# ELECTRON CLOUD SIMULATIONS IN THE FERMILAB RECYCLER*

A.P. Schreckenberger†, University of Illinois Urbana-Champaign, Urbana, IL, USA
R. Ainsworth, Fermi National Accelerator Laboratory, Batavia, IL, USA

## Abstract

We present a simulation study to characterize the stability region of the Fermilab (FNAL) Recycler in the context of secondary emission yield (SEY). Interactions between electrons and beam pipe material can produce electron clouds that jeopardize beam stability in certain focusing configurations. Such an instability was documented in the Recycler, and the work presented here reflects improvements to better understand that finding. We incorporated the Furman-Pivi Model into a PyECLOUD analysis, and we determined the instability threshold given various bunch lengths, intensities, SEY magnitudes, and model parameters.

## SECONDARY EMISSION YIELD

The secondary emission phenomenon holds a prominent position in the history of high-energy particle physics. Interactions between sufficiently-energetic incident particles and materials have the capability to liberate secondaries, and applications exploiting this effect have filled niches in scientific study—with photomultiplier tubes standing out as a dominant example. These instruments formed essential components of numerous particle physics experiments.

Conversely, the emission of secondary particles can have detrimental impacts in some scenarios. We are interested in the potential impact secondary electrons pose to the stability of the FNAL Recycler, the interactions between in-vacuum electrons and beam pipe materials, the subsequent generation of electron clouds, and a mapping of the expected stability region as a function of the SEY amplitude. The following sections of this manuscript will address the historical context of SEY investigations at FNAL and present new findings that arise from the latest simulation study.

We typically quantify the SEY strength via the $\delta$ coefficient as a function of the incident electron energy. The shapes of these functions also depend on the angle of incidence as well as the specific material surface, and they fundamentally characterize the number of electrons liberated from the impacted material by an incident particle. Analyses with an accelerator physics focus gravitate towards beam characteristics, beam pipe materials, SEY-reducing coatings, conditioning time frames, and optics-amplified instabilities. A simulated SEY curve is shown in Fig. 1, which was generated using PyECLOUD software [1] and the Furman-Pivi (FP) Model [2]. In this example, we specified inputs that describe the Recycler optics, beam intensity, and bunch structure. FP parameters, which inject surface characteristics into the electron cloud simulation, were chosen to reflect the stainless steel wall of the beam pipe.

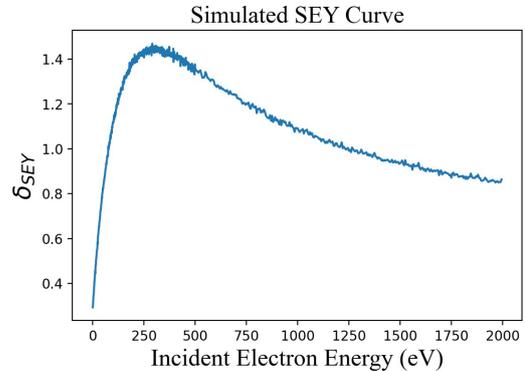

Figure 1: An SEY curve produced with PyECLOUD and the Furman-Pivi Model. Input parameters reflect conditions in the FNAL Recycler. The depicted example specifically reflects normal ($\theta$ = 0) electron incidence.

The Furman-Pivi Model considers three categories of electrons in its construction of material interactions: 1) elastic electrons, which are reflected off the surface boundary, 2) rediffused electrons, which penetrate molecular layers of the material before re-emerging, and 3) true secondaries, which are ejected from the material and can add to the number of electrons in the final state. In practice, elastic and rediffused electrons do not undergo distinct mechanisms. Both are subject to atomic scatterings that lead to a reflection of the incident particle that conserves the number of electrons in the initial and final states. The distinction in the probabilistic model stems solely from the advantages the categorization provides the phenomenological fitting process [2]. In this method, $\delta(\theta, E)$ can now be represented as the sum of its categorical components, namely:

$$\delta(\theta, E) = \delta_{ts}(\theta, E) + \delta_r(\theta, E) + \delta_e(\theta, E), \quad (1)$$

where $ts$, $r$, and $e$ refer to true secondary, rediffused, and elastic; $\theta$ is specified, and $\delta = \text{Max}(\delta(\theta, E)) = \delta_{\max}(\theta, E)$.

With the SEY curves generated to reflect the conditions in the FNAL Recycler, PyECLOUD calculates the expected electron density in the beam region as a function of time. Three examples of the simulated outputs are illustrated in Fig. 2, which demonstrates how the density behavior can change when the $\delta$ coefficient is scaled. For the dotted line in the figure with $\delta$ = 1.3, we observe that the electron density holds steady during the first 1.6 μs when proton bunches enter the accelerator and then decays over time to a final value $n_f \ll n_i$, where $f$ and $i$ designate the final and initial densities.

When $\delta$ is scaled to 1.7 and 2.0, the decay behavior after the peak persists, but the maximum density increases by orders of magnitude. Furthermore, we see a reversal in the relationship between $n_i$ and $n_f$ that potentially facilitates accelerator instabilities. When $n_f > n_i$, the beam traverses a denser electron cloud during subsequent revolutions, which occur after the time window shown in Fig. 2, and a sufficiently dense cloud could destabilize the proton beam.

___________________________________________
* Work supported by the Fermilab Accelerator Division and the University of Illinois Urbana-Champaign
† wingmc@fnal.gov, currently employed by Fermilab

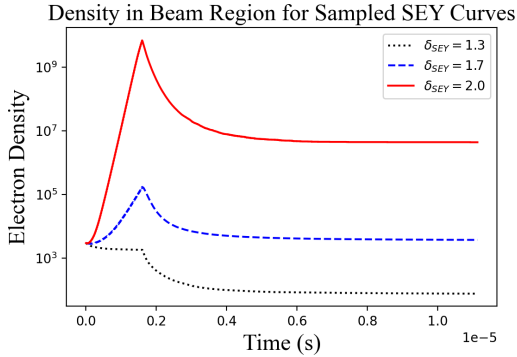

Figure 2: Simulated electron densities in the FNAL Recylcer beam region produced from a sample of SEY curves.

## FERMILAB MOTIVATION

The FNAL Recycler has a susceptibility to SEY-related instabilities because combined function magnets (CFMs) are utilized in the lattice. In 2014, J. Eldred *et al.* explored the presence of a fast transverse instability in the accelerator and attributed the cause to the buildup of an electron cloud [3]. The thesis of S. Antipov expanded upon this work by centering on several key areas, including the capability of CFMs to produce magnetic mirrors that will trap electrons, a numerical simulation of the electron trapping, and an analytical model of the cloud-driven instability. One point of tension from the Antipov thesis pertains to the range of $\delta$ that could produce the observed behavior. The simulation predicted that an electron cloud would not accumulate in the Recycler for $\delta < 2.2$ and that $\delta > 2.5$ would be required to facilitate beam-driven cloud accumulation [4].

SEY measurements conducted on stainless steel samples at FNAL, however, do not enter that high range. For an electron beam impacting a 304L planchet at a 15° angle of incidence (the empirically-determined mean incidence angle), $\delta$ ranges from $1.3-1.7$. Even under these conditions, we observed an instability in the Recycler in early 2022.

Another FNAL thesis, written by Y. Ji, revisited the instability question by exploring SEY and electron cloud simulations in the context of the FNAL Main Injector and future upgrades. POSINST simulation software inserted the FP Model into this specific study, which yielded instability thresholds with $\delta$ values consistent with SEY analyses [5]. The promising work spurred the effort presented in this manuscript by seeding two challenges. First, the range of $\delta$ reported in Ref. [5] that could facilitate instabilities extended below the minimum value from the 304L measurements. In this regard, we could hypothetically conclude that no period of accelerator operation would be deemed SEY-safe. Second, to perform this Recycler study, we needed a different simulation solution that maintained use of the FP Model and incorporated CFMs into the accelerator modeling.

## LATEST ANALYSIS

We employed the PyECLOUD+FP combination to improve the modeling of the Recycler. Ref. [2] established the stainless steel SEY parameters that we applied to the simulation, and key beam and machine inputs are detailed in Table 1. We set the nominal beam intensity to 5e10 protons-per-bunch (ppb) and the nominal bunch length to 0.4 m. The initial electron density, the nominal $n_i$, was determined in Ref. [4].

Table 1: Key Electron Cloud Simulation Inputs

| Input Name | Input Value |
|---|---|
| Beam Energy | 8.885 GeV |
| $\sigma_{x,y}$ | 0.003 m |
| Bunch Spacing | 18.936 ns |
| Nominal Intensity | $5 \times 10^{10}$ ppb |
| Nominal Bunch Length | 0.4 m |
| Nominal $n_i$ | 3000 |
| Beam Filling Profile | $84 \cdot [5 \times 10^{10}$ ppb$] + 504 \cdot [0$ ppb$]$ |

The primary objectives of this analysis included the investigation of FP integration, the creation of a stability metric, the determination of new stable $\delta$ thresholds, and the determination of SEY-stable beam intensities. Here, we introduce the ratio of the initial and final cloud densities, $R_s = n_f/n_i$, as the chosen metric. Accelerator conditions in which $R_s < 1.0$ cannot yield electron cloud accumulation scenarios. For $R_s > 1.0$, it becomes feasible for electrons to accumulate until saturation is reached, and SEY-driven instabilities are possible.

We computed $R_s$ from numerous simulations that scanned $\delta$ from $1.0-2.3$. In the standard assessment, we uniformly scaled the three categorization components ($\delta_{ts}$, $\delta_r$, $\delta_e$). We also performed a crosscheck in which $\delta_{ts}$ and $\delta_e$ were fixed, and $\delta_r$ was varied. In Figs. 3 and 4, we plot $R_s$ vs. $\delta_{max}$, which is the maximum $\delta$ from the simulated SEY curve given a 15° incidence angle, $\delta_{max} = \delta_{max}(15°, E)$.

Figure 3 displays an FP Model systematic assessment, which gauged the importance of the SEY fit inputs. Each point represents a single simulation in the study, and the dashed vertical lines depict the data range of measured $\delta$ for a 304L Stainless Steel beam pipe. Both the uniform-scaling and $\delta_r$-targeted scans of the SEY amplitude are pictured. However, the focus of this plot is the impact that ±25% shifts on the FP shape parameters have on $R_s$. The magnitudes of these shifts are significantly larger than any real-world material aberration. In some cases, a 25% change exceeds what a complete material swap would induce (e.g. the true-secondary shape parameter, s, for stainless steel vs. copper).

Given that only the $s$ and $\hat{E}_0$ shifts induced changes in $R_s$ that crossed the 1.0 stability boundary, while those margins are representative of full material replacements as opposed to minor surface anomalies, we built confidence that conclusions derived from beam parameter scans would not be swayed by potential divergences from the SEY fit results provided by Ref. [2].

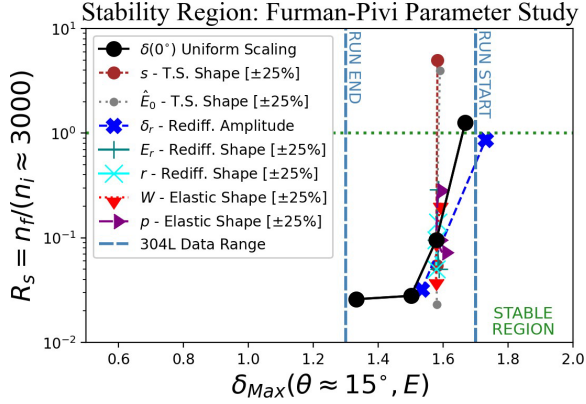

Figure 3: $R_s$ vs. $\delta_{max}$ given shifts in the FP parameters. The uniform-scaling (black points, solid line) and $\delta_r$-targeted (bold blue xs, dashed line) scans of the SEY amplitude are shown. Key parameters, which define the shape of the SEY curve, were shifted by $\pm 25\%$ and are labeled in the legend. The range of SEY amplitudes measured in the FNAL Main Injector using an electron beam incident on 304L Stainless Steel is depicted by the two dashed vertical lines.

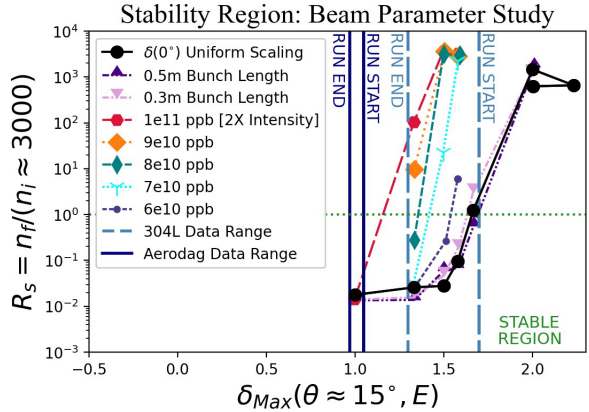

Figure 4: $R_s$ vs. $\delta_{max}$ for beam parameter simulations. The uniform-scaling scan (black points, solid line) of the SEY amplitude is shown over the full $1.0-2.3$ range along with $\pm 0.1$ m shifts in the bunch length (dash-dotted lines, triangular markers). The beam intensity was assessed from $5 \times 10^{10} - 1 \times 10^{11}$ ppb, and markers are labeled in the legend. The dotted horizontal line designates the $R_s = 1.0$ stabilty threshold. The dashed vertical lines denote the FNAL Main Injector 304L Stainless Steel data range, and the vertical solid lines give the data range for a beam pipe coated in Aerodag.

The results shown in Fig. 4 provide several important pieces of information. The dashed vertical lines indicate the range of SEY measurements taken during the 2021 Run with a 304L Stainless Steel beam pipe, and the solid vertical lines depict measurements taken during the same time period with an Aerodag-treated sample (a graphite-based coating). We also observed an asymptotic reduction in the $\delta$ measurements over the course of the run, reaffirming the surface conditioning processes described in Refs. [4, 5].

From the scan of the FP $\delta$ parameters, we find a convergence with the data measurements, proximity to the stability threshold, and conditioning process. The 2022 instability came after a 12-day downtime, which is sufficient for deconditioning. The beam had also been tuned to a bunch length of 0.3 m. In this case, we expect $\delta_{max} = 1.7$, and the associated simulations exceed $R_s = 1.0$ when considering $0.3$ m$-0.5$ m bunch lengths. The simulation suggests that conditions in that moment do facilitate an instability. However, after sufficient conditioning occurs, $\delta_{max}$ will decrease, $R_s \rightarrow 10^{-2}$, and the Recycler becomes stable. This simulation settles the $\delta$ range question from Ref. [4] and follows the observed behavior of the accelerator.

We also determined the maximum intensity the Recycler can circulate. With an untreated beam pipe, we could reach $8 \times 10^{10}$ ppb after a sufficiently long conditioning period. Considering the intent of the FNAL PIP-II upgrade to increase the intensity by 50%, this study demonstrates that this objective encroaches upon the limit of the accelerator once one accounts for realistic maintenance downtime.

## CONCLUSION

We deployed a new simulation to assess the impact of SEY on the stability of the FNAL Recycler and tested the FP implementation. Improving upon previous analyses, the new stability metric, $R_s$, appears to accurately gauge when accelerator conditions facilitate destabilizing electron cloud buildups, and we placed a tentative limit of $8 \times 10^{10}$ ppb on the beam intensity that the Recycler could theoretically circulate.

## ACKNOWLEDGMENTS

Special thanks to Giovanni Iadarola for his assistance with debugging and developing the simulation software.